\documentstyle[prb,aps]{revtex}
\draft
\begin{document}
\def\SNG{{\em Physical Review Style and Notation Guide}}
\def\LUG {{\em \LaTeX{} User's Guide \& Reference Manual}}
\def\btt#1{{\tt$\backslash$\string#1}}%
\def\REVTeX{REV\TeX}
\def\AmS{{\protect\the\textfont2
        A\kern-.1667em\lower.5ex\hbox{M}\kern-.125emS}}
\def\AmSLaTeX{\AmS-\LaTeX}
\def\BibTeX{\rm B{\sc ib}\TeX}
\twocolumn[\hsize\textwidth\columnwidth\hsize\csname@twocolumnfalse%
\endcsname
\title{The Eliashberg function of amorphous metals}
\author{D.Belitz and M.N.Wybourne}
\address{Department of Physics and Materials Science Institute,\\
 University of Oregon,\\
 Eugene, OR 97403}

\date{\today}
\maketitle
\begin{abstract}
A connection is proposed between the anomalous thermal transport
properties of amorphous solids and the low-frequency behavior of
the Eliashberg function. By means of a model calculation we show
that the size and frequency dependence of the phonon mean-free-path
that has been extracted from measurements of the thermal conductivity
in amorphous solids leads to a sizeable linear region in the
Eliashberg function at small frequencies. Quantitative comparison
with recent experiments gives very good agreement.
\end{abstract}

]
\narrowtext

The functional form of the Eliashberg function, $\alpha^2F(\omega)$,
in amorphous metals
has been the subject of debate for a long time. Experimentally, there
is uniform agreement that in amorphous simple metals the low-frequency
part of the Eliashberg function is strongly enhanced over what is
observed in crystalline materials.\cite{BergmannR}
The functional form at low frequencies,
although difficult to determine experimentally, is usually found to
be linear.\cite{transitionmetals}
This experimental observation has given rise to a
substantial theoretical debate. Bergmann\cite{Bergmann71} and
many others\cite{linear} argued that a disorder enhancement of the
electron-phonon coupling leads to a linear low-frequency behavior of
the Eliashberg function, in accord with experiment.
This was disputed by Schmid \cite{Schmid73}
and by Keck and Schmid\cite{KeckSchmid}, who claimed that this
disorder enhancement was spurious, and due to an incorrect application
of the Fr\"ohlich model to disordered materials.
Schmid's calculation gave instead
$\alpha^2F(\omega\rightarrow 0)\sim\omega^3$, in disagreement with experiment.
The $\omega^3$ result was also confirmed by
others.\cite{Eisenriegler} Since different models and different
calculational methods had been used by these various authors, the
theoretical problem was widely considered unsolved for a long time.
This disagreement was finally
settled by Reizer and Sergeyev,\cite{ReizerSerg}
who explicitly pointed out the errors in Ref.\ \onlinecite{linear}
and showed that a correct calculation leads to Schmid's result
independent of the model and the method used. This led to the
unsatisfying situation that a theoretically credible result, viz.
Schmid's $\alpha^2F(\omega\rightarrow 0)\sim\omega^3$,
disagreed with experiment,
while incorrect arguments led to
$\alpha^2F(\omega\rightarrow 0)\sim\omega$
in agreement with the experimental observations.
This situation was summarized by one of us,\cite{DB} who also argued that
no existing comparison between theory and experiment had been careful
and accurate enough to really rule out Schmid's result. However,
recent experiments by Watson and Naugle have
shown that Schmid's result is not compatible with experiments
on amorphous CuSn alloys.\cite{Naugle}

At this point it is important to distinguish between Schmid's general
theory for the electron-phonon coupling in impure metals, and his and
others' specific model calculations. Since the work of Reizer and
Sergeyev showed that Schmid's general theory is physically
correct, the search for reasons behind the discrepancy between his
results and experiment should then turn to the model
assumptions. On the electronic
side the main assumption is that of nearly free electrons. It is hard
to see how this could qualitatively fail in simple metals as long as
the resistivities are moderate, and the electronic states are effectively
three-dimensional. On the
phonon side, Schmid assumed undamped Debye phonons, an assumption
that is necessary in order to obtain the $\omega^3$-law as the
asymptotic low-frequency behavior. As was shown in Ref.\ \onlinecite{DB},
the inclusion of phonon damping by electrons leads to a {\it linear}
low-frequency asymptotic behavior (albeit with a prefactor that is too
small by several orders of magnitude to explain the experimental results),
which then crosses over to Schmid's $\omega^3$-law. The prefactor of
the linear term is proportional to the phonon damping. This raises the
possibility that very strong phonon damping (which would have to be of
other than electronic origin) might lead to a linear term in
$\alpha^2F(\omega)$ whose prefactor is large enough to account for the
experimental observations.

In order to pursue this last point, let us recall that besides the
problems with the Eliashberg function mentioned above, amorphous
materials have properties of entirely phononic origin that
are hard to understand. In particular the thermal
conductivity, $\kappa$, shows
an enigmatic behavior. Even though it has been stressed that the
thermal conductivity is not understood in any temperature region,
the general phenomenology is clear, consistent, and well
documented.\cite{FreemanAnderson,amorphoussolids}
As a function of temperature $T$, the
thermal conductivity behaves like $\kappa\sim T^2$ for $T/\Theta\alt 10^{-2}$,
with $\Theta$ the Debye temperature.
The origin for the phonon scattering in this region is not known for
certain. The phenomenological two-level system concept has often been
invoked in this context,\cite{amorphoussolids}
but no consensus has ever been reached. For
$10^{-2}\alt T/\Theta\alt 10^{-1}$ the thermal conductivity is
approximately independent of $T$. This is the so-called plateau region,
which is characterized by strong, and strongly frequency dependent,
phonon scattering of uncertain origin. Finally, for $T/\Theta\agt 10^{-1}$
the thermal conductivity becomes $T$-dependent again, but it is not
even clear whether the heat transport in this region is by phonons, much
less what the scattering mechanisms are.

This poor state of physical understanding notwithstanding, the above
phenomenology is remarkably universal, and seems to be characteristic
of amorphous materials, both insulating,\cite{FreemanAnderson}
and metallic.\cite{metals} It has been used to
deduce the following behavior of the phonon mean-free path, $l_{ph}$, as a
function of frequency. For frequencies
$\omega\alt 10^{-2} k_B\Theta/\hbar$, $l_{ph}$ is a linear function of
frequency. For intermediate frequencies, $10^{-2}\alt\hbar\omega/k_B\Theta
\alt 10^{-1}$, $l_{ph}$ goes as a high power, $n$, of frequency. $n$ has
been reported to be at least $4$, and possibly larger. This intermediate
frequency regime corresponds to the plateau region in the thermal
conductivity. At still higher frequencies,
$\omega\agt 10^{-1} k_B\Theta/\hbar$, the phonon mean free path either
becomes frequency independent,\cite{ACAnderson} or is a linear function
of frequency again.\cite{FreemanAnderson}

In this paper we propose a connection between the thermal properties
of amorphous materials as described above, and the
low-frequency behavior of the Eliashberg
function. In particular we show that Anderson's
phenomenological functional form of the phonon
mean-free path, if used in Schmid's
theory for the electron-phonon coupling, explains the observed
behavior of the Eliashberg function as well as the observed behavior
of the thermal transport. Let us start from the expression
for the Eliashberg function, based on Schmid's general theory,\cite{Schmid73}
that was derived in Ref.\ \onlinecite{DB},
\begin{mathletters}
\label{eqs:1}
\begin{equation}
\alpha^2F(\omega) = \frac{1}{2\pi^2 N_F}\sum_{{\bf q},b}\ \alpha_b({\bf q})\
   \frac{c_b}{\omega_b^2({\bf q})}\ {\rm Im}\,D_b^R({\bf q},\omega)\quad.
\label{eq:1a}
\end{equation}
Here $D_b^R({\bf q},\omega)$ is the retarded phonon propagator, whose
imaginary part reads,
\begin{equation}
{\rm Im}\,D_b^R({\bf q},\omega) =
                   \frac{4\omega\omega_b^2({\bf q})\gamma_b({\bf q})}
                        {\left(\omega^2 - \omega_b^2({\bf q})\right)^2
                                              + 4\omega^2\gamma_b^2({\bf q})}
\quad,
\label{eq:1b}
\end{equation}
\end{mathletters}%
where $\gamma({\bf q})$ is the phonon damping coefficient. In writing
Eqs.\ (\ref{eqs:1}) we have assumed a free electron model with $N_F$ the
electronic density of states per spin at the Fermi level. We have also
assumed Debye phonons with one longitudinal and two transverse branches
labeled by $b$ ($b=L,T$), speed of sound $c_b$, and dispersion
$\omega_b({\bf q}) = c_b q$. $\alpha_b({\bf q})$ in Eq.\ (\ref{eq:1a})
is the electronic contribution to the sound attenuation coefficient,
for which we use the standard Pippard result,\cite{Pippard}
\begin{mathletters}
\label{eqs:2}
\begin{equation}
\alpha_b({\bf q}) = \kappa_b\,f_b(ql)\quad,
\label{eq:2a}
\end{equation}
where $\kappa_b = (v_F/c_b)(\rho_e/\rho_{ion})/l$ with $v_F$ the Fermi
velocity, $l$ the electronic mean-free path, and $\rho_e$ and
$\rho_{ion}$ the electronic and ionic mass density, respectively. The
functions $f_{L,T}$ are given by,\cite{Pippard}
\begin{equation}
f_L(x) = \frac{1}{3}\ \frac{x^2\arctan(x)}{x-\arctan(x)}\ -\ 1\quad,
\label{eq:2b}
\end{equation}
\begin{equation}
f_T(x) = \frac{1}{2x^3}\ \left[2x^3 + 3x - 3(x^2+1)\arctan(x)\right]\quad.
\label{eq:2c}
\end{equation}
\end{mathletters}%
With phonon damping exclusively by electrons, as was assumed in Ref.\
\onlinecite{DB}, one has $\gamma_b({\bf q}) = c_b\alpha_b({\bf q})/2$.
Here, however, we will consider the possibility of nonelectronic
contributions to $\gamma_b({\bf q})$. Accordingly, we write
\begin{equation}
\gamma_b({\bf q}) = \tilde\gamma\,c_b\,q_D\,g(q/q_D)\quad,
\label{eq:3}
\end{equation}
where $q_D$ is the Debye wavenumber, $\tilde\gamma$ is a number,
and $g$ is some function that determines the wavenumber or frequency
dependence of the phonon damping. The latter we model after Anderson's
proposal,\cite{ACAnderson} which has been extracted phenomenologically
from thermal transport measurements in amorphous materials.
Anderson's model consists of the three distinct regions mentioned above:
(1) A low frequency region
where the damping is a linear function of frequency, (2) an intermediate
region where the damping goes as a large power of the frequency, and
(3) a high frequency region where the damping is independent of
frequency.\cite{highfrequencyfootnote}
The intermediate region corresponds to the characteristic plateau that
is observed in the T-dependent thermal conductivity. We thus model the
function $g(x)$ in Eq.\ (\ref{eq:3}) as,
\begin{equation}
g(x) = 10^n y\ \frac{x/y + (x/y)^n}{10^n + (x/y)^n}\quad.
\label{eq:4}
\end{equation}
Here $y$ is the onset of the plateau region in units of the Debye
wavenumber, the width of the plateau region has been assumed to be one
decade, and $n$ is the power that characterizes the frequency dependence
of the phonon mean-free path in the plateau region.

Before we turn to a numerical evaluation of the integral, Eq.\ (\ref{eq:1a}),
that determines $\alpha^2F(\omega)$, let us consider the low-frequency
behavior analytically. Asymptotically,
$\alpha^2F(\omega)\sim\omega/\omega_{\alpha}$,
with a slope $\omega_{\alpha}^{-1}$.
The latter we estimate for a clean system,
i.e. in the limit $l\rightarrow\infty$. In this limit only longitudinal
phonons contribute, and we can use the asymptotic form of the function
$f_L$ in Eq.\ (\ref{eq:2b}), $f_L(x\rightarrow\infty) = \pi x/6$. Then we
obtain,
\begin{equation}
\frac{\epsilon_F}{\omega_{\alpha}} = \frac{\tilde\gamma}{6\pi}\
     \frac{q_D}{k_F}\ \left(\frac{v_F}{c_L}\right)^3\
     \frac{\rho_e}{\rho_{ion}}\
     \int_0^1 \frac{dx}{x}\ g(x)\quad.
\label{eq:5}
\end{equation}
Typical parameter values are $q_D/k_F\approx 1$, $v_F/c_L\approx 10^3$, and
$\rho_e/\rho_{ion}\approx 10^{-5}$. For the parameters $y$ and $n$ in
Eq.\ (\ref{eq:4}) we take \cite{ACAnderson} $y\approx 0.02$
and $n\approx 4$. Finally, $\tilde\gamma$ determines the overall scale
for the phonon mean-free path $l_{ph}$. A typical value is
$l_{ph}\approx 1{\rm cm}$ at a frequency of 1GHz. With
$c_L\approx 2\times 10^5 {\rm cm/s}$ this corresponds to
$\tilde\gamma\approx 2\times 10^{-4}$. This yields
$\epsilon_F/\omega_{\alpha}\approx 1,700$. With a Fermi energy
$\epsilon_F\approx 10{\rm eV}$ we obtain
$\omega_{\alpha}\approx 6{\rm meV}$. This value for $\omega_{\alpha}$ is
of the same order of magnitude as the one
typically obtained from tunneling experiments.\cite{BergmannR,Naugle}

Now that we have seen that we obtain promising results for
$\alpha^2F(\omega\rightarrow 0)$ with reasonable parameter values,
let us calculate $\alpha^2F(\omega)$ numerically, and compare
quantitatively with experiments. Watson and Naugle\cite{Naugle} have
performed a detailed study of amorphous SnCu. For the stoichiometry
${\rm Sn}_{.87} {\rm Cu}_{.13}$ they quote
the following parameter values:
$\epsilon_F = 1.54\times 10^{-11}{\rm erg}$,
$k_F = 1.59\times 10^8{\rm cm^{-1}}$,
$v_F = 1.84\times 10^8{\rm cm/s}$,
$l = 9.58\times 10^{-9}{\rm cm}$,
$q_D = 1.31\times 10^8{\rm cm^{-1}}$,
$c_L =1.6\times 10^5{\rm cm/s}$, $c_T = 8.1\times 10^4{\rm cm/s}$.
Of these, the electronic
parameters are much better known than the two sound velocities.
Using these parameters, as well as $n=5$, $y=0.015$, and $\tilde\gamma=
8.0\times 10^{-5}$,
we have calculated $\alpha^2F(\omega)$
for frequencies up to
$1.6{\rm meV}$, which was the lower frequency cutoff in the
experiment of Ref.\ \onlinecite{Naugle}. The high frequency behavior
resulting from our calculation would not be realistic anyway due to
our using a Debye model. The result was shown in Ref.\ \onlinecite{Naugle},
and was used as low-frequency input in a McMillan-Rowell inversion
procedure to obtain $\alpha^2F$ from tunneling data.
It is also shown again as the curve
labeled $n=5$ in Fig.\ \ref{fig:1}. For the inversion
procedure an overall factor multiplying the calculated $\alpha^2F$ was
used as a fit parameter. The need for such an overall scale factor is
not surprising, given our free electron model. The factor used for the
best fit is equivalent to a deviation of the density of states in
Eq.\ (\ref{eq:1a}) from its free electron value by 14\%.
A comparison
between the calculated and the measured tunneling density of states
then provides a measure of how well the low-frequency input describes
the actual system. Watson and Naugle found that our calculated $\alpha^2F$
does very well, although not quite as well as if one assumes a strictly
linear low-frequency behavior. It should be stressed that our
calculation used strictly the parameters as provided by the
experimentalists, some of which are not known very accurately. Since
the inversion procedure is quite involved no attempt was made to
fine tune the parameters.

This result shows that Schmid's theory with
a phonon damping that accounts for the thermal transport properties
characteristic of amorphous metals gives good agreement between the
calculated Eliashberg function and tunneling data. In contrast, the
same theory with phonon damping by electrons only is not capable of
explaining the experimental results.\cite{Naugle}

In addition to this comparison between theory and experiment, let
us demonstrate the effects of some parameter changes on $\alpha^2F$.
We consider the four results for $\alpha^2F$ shown in Fig.\ \ref{fig:1}.
The curve labeled $n=5$ was obtained with the parameters as given above.
The slight bulge in this curve results from the leveling off of the
phonon mean-free path at the high-frequency end of the plateau region.
This moderates
the rapid increase of $\alpha^2F$ at lower frequencies, which is due
to the strong frequency dependence of the phonon mean-free path. With
a weaker frequency dependence of the phonon mean-free path in the plateau
region, i.e. a smaller exponent $n$ in Eq.\ (\ref{eq:4}), the initial
slope of $\alpha^2F$ is much smaller, and over the frequency range
considered $\alpha^2F$ shows a purely positive curvature. Conversely,
a still larger exponent $n$ leads to a purely negative
curvature of $\alpha^2F$. This is demonstrated in Fig.\ \ref{fig:1}.
The curves with stronger curvature all led to substantially less good
agreement with experiment than the one for $n=5$.
We have also considered the sensitivity of the result to the ratio
of the longitudinal and transverse speeds of sound, which is not
known very accurately. We have found only a very weak dependence of
the functional form of $\alpha^2 F$ on this ratio in the region
$1.8<c_L/c_T<2.5$. Finally, we have changed the damping parameter,
$\tilde\gamma$, with all other parameters held fixed. This was found
to have a very similar effect to changing $n$, with $\alpha^2 F$ changing
from negative to positive curvature as
$\tilde\gamma$ is increased, or the phonon mean-free path at a reference
frequency is decreased. The effect of changing $\tilde\gamma$ by a
factor of ten was roughly equivalent to changing $n$ by one.
For instance, with $n=4$ and $\tilde\gamma = 8\times 10^{-4}$ we
obtained a curve that was hardly distinguishable from the one
for $n=5$ shown in Fig.\ \ref{fig:1}. Generally, we found that with
reasonable parameters for simple metals we need $4\alt n\alt 6$ in order
for our explanation of the behavior of $\alpha^2 F$ to be viable.

In conclusion, we have shown that Schmid's theory of electron-phonon
coupling in impure metals can account for the observed low-frequency
behavior of the Eliashberg function in amorphous simple metals if one
assumes a strong phonon damping consistent with the one extracted from
measurements of the thermal conductivity. While the physics underlying
the strong damping is not known, this observation unifies two seemingly
unconnected, and separately mysterious, properties of
amorphous materials. It suggests that strong phonon scattering is a
very fundamental feature of the amorphous state, and that understanding
its origin would explain many different properties of amorphous materials
at once.

We would like to thank Don Naugle for rekindling our interest in
the present problem, and for numerous discussions.
This work was supported by the NSF under
grant numbers DMR-92-09879 and DMR-90-19525.

\begin{figure}
\caption{Results for the Eliashberg function $\alpha^2F$ as a function
 of energy or frequency for different values of the exponent $n$ in
 Eq.\ (\protect\ref{eq:4}). All other parameters were held fixed at the
 values given in the text. $\alpha^2F$ has been normalized by its value
 at $\omega=1.6{\rm meV}$, which is $0.436$, $0.279$, $0.208$, and $0.200$
 for $n=6$, $5$, $4$, and $3$, respectively.}
\label{fig:1}
\end{figure}


\begin{references}
\bibitem{BergmannR} See, e.g., G. Bergmann, Phys. Rep. {\bf 27C}, 159 (1976).
\bibitem{transitionmetals} It is worth mentioning that the low-frequency
 enhancement may be absent, and the low-frequency behavior of
 $\alpha^2F$ smaller than linear, in amorphous transition metals,
 D.~B. Kimhi and T.~H. Geballe, Phys. Rev. Lett. {\bf 45}, 1039 (1980).
 This difference between simple metals and transition metals is not well
 understood, and needs more experimental attention. For the purpose of
 our discussion we restrict ourselves to simple metals.
\bibitem{Bergmann71} G. Bergmann, Phys. Rev. B {\bf 3}, 3797 (1971).
\bibitem{linear} H. Takayama, Z. Phys. {\bf 263}, 329 (1973);  S.G. Lisitin,
 Fiz. Nizk. Temp. {\bf 1}, 1516 (1975) [Sov. J. Low Temp. Phys. {\bf 1}, 728
 (1975)]; S.J. Poon and T.H. Geballe, Phys. Rev. B {\bf 18}, 233 (1978).
\bibitem{Schmid73} A. Schmid, Z. Phys. {\bf 259}, 421 (1973).
\bibitem{KeckSchmid} B. Keck and A. Schmid, J. Low Temp. Phys. {\bf 24}, 611
 (1976).
\bibitem{Eisenriegler} E. Eisenriegler, Z. Phys. {\bf 258}, 185 (1973);
 G. Gr\"unewald and K. Scharnberg, Z. Phys. {\bf 268}, 197 (1974); Z. Phys.
 B {\bf 20}, 61 (1975).
\bibitem{ReizerSerg} M.Yu. Reizer and A.V. Sergeyev, Zh. Eksp. Teor. Fiz.
 {\bf 90}, 1056 (1986) [Sov. Phys. JETP {\bf 63}, 616 (1986)].
\bibitem{DB} D. Belitz, Phys. Rev. B {\bf 36}, 2513 (1987).
\bibitem{Naugle} P.~W. Watson III and D.~G. Naugle, preceding paper,
 Phys. Rev. B {\bf xx}, xxx (1995); and private communication.
\bibitem{FreemanAnderson} J.~J. Freeman and A.~C. Anderson,
 Phys. Rev. B {\bf 34}, 5684 (1986).
\bibitem{amorphoussolids} {\it Amorphous Solids}, edited by
 W.~A. Phillips, Springer (New York 1981).
\bibitem{metals} H.~v. L\"ohneysen and F. Steglich, Phys. Rev. Lett.
 {\bf 39}, 1205 (1977); J.~R. Matey and A.~C. Anderson, J. Non-Cryst.
 Solids {\bf 23}, 129 (1977).
\bibitem{ACAnderson} A.~C. Anderson, in {\it Amorphous Solids},
 Ref.\ \onlinecite{amorphoussolids}, Fig.5.3.
\bibitem{Pippard} A.~B. Pippard, Philos. Mag. {\bf 46}, 1104 (1955).
\bibitem{highfrequencyfootnote} As mentioned before, the behavior in
 the high-frequency region, above the plateau, is uncertain. We have
 also tried a linear frequency dependence for $l_{ph}$ in this
 region.\cite{FreemanAnderson} In our results for $\alpha^2F$ we have
 found no qualitative differences between this assumption and the one
 of a frequency independent $l_{ph}$.
\end{references}
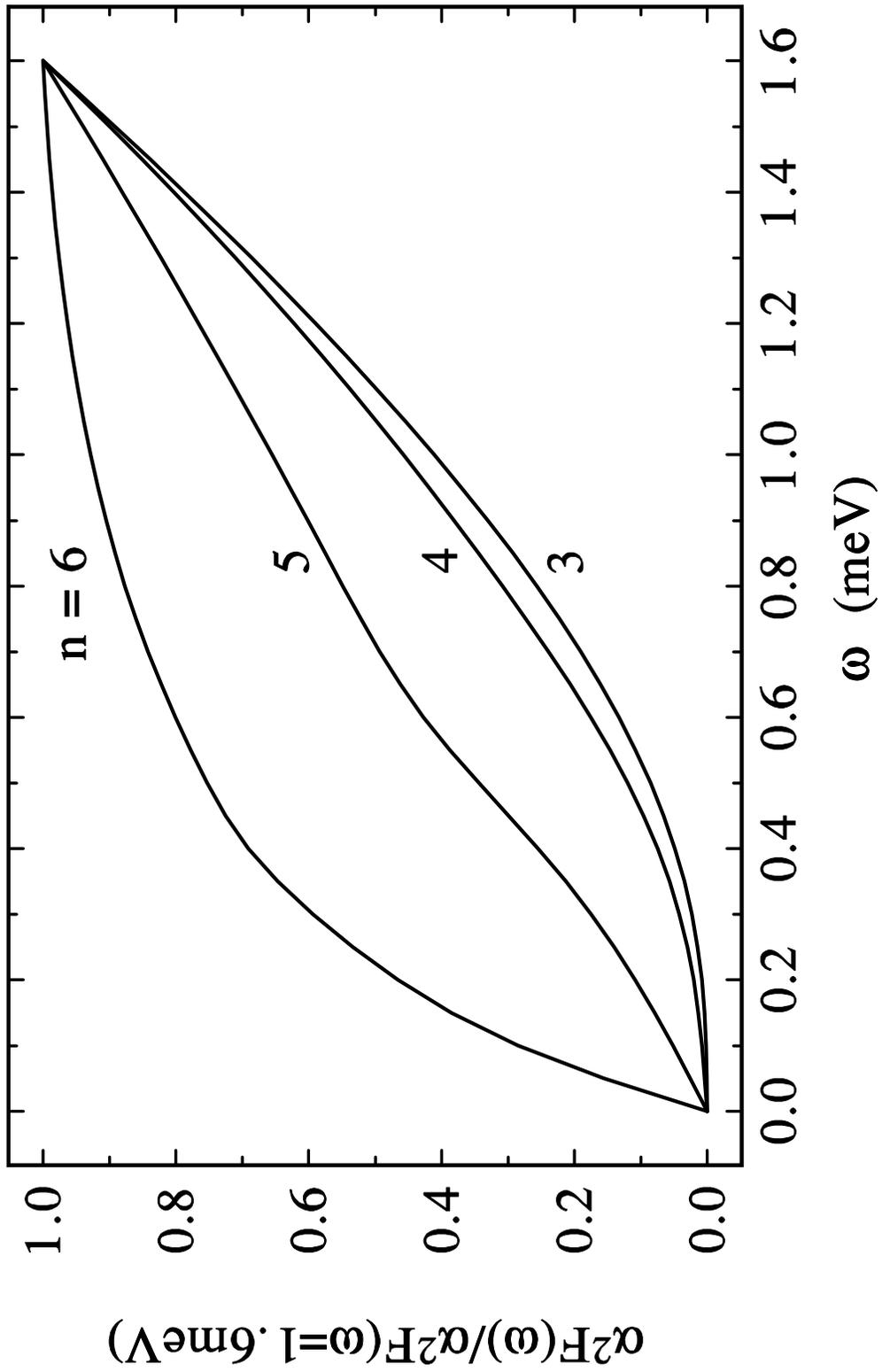
\end{document}